\documentclass[aps,prl,preprint,tightenlines,superscriptaddress,showpacs,byrevtex]{revtex4}
\usepackage{graphicx} 
\usepackage{dcolumn}  

\graphicspath{{ps}}

\def\Journal#1#2#3#4{{#1} {\bf #2}, #3 (#4)}

\def\NIMA{Nucl. Instr. and Meth. A}

\def\PLB{{Phys. Lett.}  B}
\def\PRL{Phys. Rev. Lett.}
\def\PRD{{Phys. Rev.} D}

\def\be{\begin{equation}}
\def\ee{\end{equation}}
\def\bea{\begin{eqnarray}}
\def\eea{\end{eqnarray}}

\begin{document}


\preprint{\vbox{ \hbox{   }
                 \hbox{BELLE-CONF-0444}
                 \hbox{ICHEP04 8-0691} 
}}

\title{\quad\\[0.5cm]  \boldmath Study of the Suppressed Decays $B^{\pm} \to[K^{\mp}\pi^{\pm}]_{D}K^{\pm}$ at Belle}

\affiliation{Aomori University, Aomori}
\affiliation{Budker Institute of Nuclear Physics, Novosibirsk}
\affiliation{Chiba University, Chiba}
\affiliation{Chonnam National University, Kwangju}
\affiliation{Chuo University, Tokyo}
\affiliation{University of Cincinnati, Cincinnati, Ohio 45221}
\affiliation{University of Frankfurt, Frankfurt}
\affiliation{Gyeongsang National University, Chinju}
\affiliation{University of Hawaii, Honolulu, Hawaii 96822}
\affiliation{High Energy Accelerator Research Organization (KEK), Tsukuba}
\affiliation{Hiroshima Institute of Technology, Hiroshima}
\affiliation{Institute of High Energy Physics, Chinese Academy of Sciences, Beijing}
\affiliation{Institute of High Energy Physics, Vienna}
\affiliation{Institute for Theoretical and Experimental Physics, Moscow}
\affiliation{J. Stefan Institute, Ljubljana}
\affiliation{Kanagawa University, Yokohama}
\affiliation{Korea University, Seoul}
\affiliation{Kyoto University, Kyoto}
\affiliation{Kyungpook National University, Taegu}
\affiliation{Swiss Federal Institute of Technology of Lausanne, EPFL, Lausanne}
\affiliation{University of Ljubljana, Ljubljana}
\affiliation{University of Maribor, Maribor}
\affiliation{University of Melbourne, Victoria}
\affiliation{Nagoya University, Nagoya}
\affiliation{Nara Women's University, Nara}
\affiliation{National Central University, Chung-li}
\affiliation{National Kaohsiung Normal University, Kaohsiung}
\affiliation{National United University, Miao Li}
\affiliation{Department of Physics, National Taiwan University, Taipei}
\affiliation{H. Niewodniczanski Institute of Nuclear Physics, Krakow}
\affiliation{Nihon Dental College, Niigata}
\affiliation{Niigata University, Niigata}
\affiliation{Osaka City University, Osaka}
\affiliation{Osaka University, Osaka}
\affiliation{Panjab University, Chandigarh}
\affiliation{Peking University, Beijing}
\affiliation{Princeton University, Princeton, New Jersey 08545}
\affiliation{RIKEN BNL Research Center, Upton, New York 11973}
\affiliation{Saga University, Saga}
\affiliation{University of Science and Technology of China, Hefei}
\affiliation{Seoul National University, Seoul}
\affiliation{Sungkyunkwan University, Suwon}
\affiliation{University of Sydney, Sydney NSW}
\affiliation{Tata Institute of Fundamental Research, Bombay}
\affiliation{Toho University, Funabashi}
\affiliation{Tohoku Gakuin University, Tagajo}
\affiliation{Tohoku University, Sendai}
\affiliation{Department of Physics, University of Tokyo, Tokyo}
\affiliation{Tokyo Institute of Technology, Tokyo}
\affiliation{Tokyo Metropolitan University, Tokyo}
\affiliation{Tokyo University of Agriculture and Technology, Tokyo}
\affiliation{Toyama National College of Maritime Technology, Toyama}
\affiliation{University of Tsukuba, Tsukuba}
\affiliation{Utkal University, Bhubaneswer}
\affiliation{Virginia Polytechnic Institute and State University, Blacksburg, Virginia 24061}
\affiliation{Yonsei University, Seoul}
  \author{K.~Abe}\affiliation{High Energy Accelerator Research Organization (KEK), Tsukuba} 
  \author{K.~Abe}\affiliation{Tohoku Gakuin University, Tagajo} 
  \author{N.~Abe}\affiliation{Tokyo Institute of Technology, Tokyo} 
  \author{I.~Adachi}\affiliation{High Energy Accelerator Research Organization (KEK), Tsukuba} 
  \author{H.~Aihara}\affiliation{Department of Physics, University of Tokyo, Tokyo} 
  \author{M.~Akatsu}\affiliation{Nagoya University, Nagoya} 
  \author{Y.~Asano}\affiliation{University of Tsukuba, Tsukuba} 
  \author{T.~Aso}\affiliation{Toyama National College of Maritime Technology, Toyama} 
  \author{V.~Aulchenko}\affiliation{Budker Institute of Nuclear Physics, Novosibirsk} 
  \author{T.~Aushev}\affiliation{Institute for Theoretical and Experimental Physics, Moscow} 
% ...
\collaboration{The Belle Collaboration}

\noaffiliation

\begin{abstract}
We report a study of the suppressed decay $B^{-} \to [K^{+}\pi^{-}]_{D}K^{-}$(and its charge-conjugate mode) at Belle, where $[K^{+}\pi^{-}]_{D}$ indicates that the $K^{+}\pi^{-}$ pair originates from a neutral $D$ meson. A data sample containing 274 million $B\bar{B}$ pairs recorded at the $\Upsilon(4S)$ resonance with the Belle detector at the KEKB asymmetric $e^{+}e^{-}$ storage ring is used. This decay mode can be used to extract the CKM angle $\phi_{3}$ using the so-called Atwood-Dunietz-Soni method. The signal for $B^{-} \to [K^{+}\pi^{-}]_{D}K^{-}$ has $2.7\sigma$ statistical significance, and we set a limit on the ratio of $B$ decay amplitudes $r_B < 0.28$ at the $90\%$ confidence level. We observe a signal with $5.8\sigma$ statistical significance in the related mode, $B^{-} \to [K^{+}\pi^{-}]_{D}\pi^{-}$.
\end{abstract}


\maketitle

\tighten

{\renewcommand{\thefootnote}{\fnsymbol{footnote}}}
\setcounter{footnote}{0}

\section{\hspace{0.38\textwidth} \boldmath Introduction}
\begin{figure}[h]
\begin{center}
\includegraphics[width=0.5\textwidth]{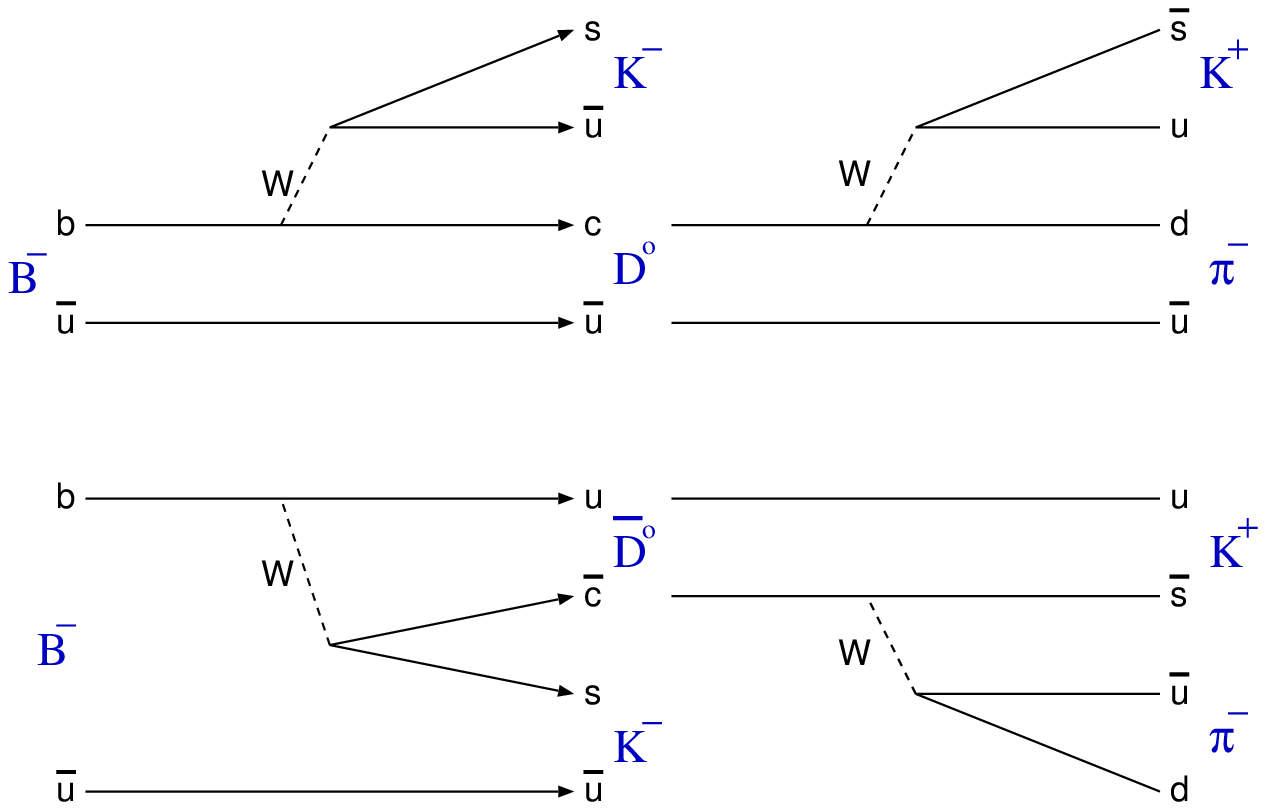}
\caption{$B^{-}\to[K^{+}\pi^{-}]_{D}K^{-}$ decays.}
\label{fig:btodcsk}
\end{center}
\end{figure}

The extraction of $\phi_{3}$, an angle in the Kobayashi-Maskawa triangle\cite{km}, is a challenging measurement even with modern high luminosity $B$ factories. Several methods for measuring $\phi_{3}$ use the interference between $B^{-} \to D^{0}K^{-}$ and $B^{-} \to \bar{D}^{0}K^{-}$, which occurs when $D^{0}$ and $\bar{D}^{0}$ decay to common final states\cite{gw}. In this paper, we analyze the suppressed decay $B^{-}\to[K^{+}\pi^{-}]_{D}K^{-}$ and its charge conjugate mode, where $[K^{+}\pi^{-}]_{D}$ indicates that the $K^{+}\pi^{-}$ pair originates from a neutral $D$ meson. In this case, the color-allowed $B$ decay followed by the doubly Cabbibo-suppressed $D$ decay interferes with the color-suppressed $B$ decay followed by the Cabbibo-allowed $D$ decay(Fig.\ref{fig:btodcsk}). This decay mode can be used to extract $\phi_{3}$ using the so-called Atwood-Dunietz-Soni method(ADS method)\cite{ads}.

{\begin{center}
\section{\boldmath ADS method}
\end{center}}
Here we define the amplitudes for $B$ decays and $D$ decays as follows:
\begin{eqnarray}
A_{B} &\equiv& A(B^{-} \to D^{0}K^{-}) = A(B^{+} \to \bar{D}^{0}K^{+}), \:\:\:\: A_{D} \equiv A(D^{0} \to \bar{f}) = A(\bar{D}^{0} \to f), \nonumber \\
\bar{A}_{B} &\equiv& A(B^{-} \to \bar{D}^{0}K^{-}) = A(B^{+} \to {D}^{0}K^{+}), \:\:\:\: \bar{A}_{D} \equiv A(D^{0} \to {f}) = A(\bar{D}^{0} \to \bar{f}). \nonumber
\end{eqnarray}
The branching fractions for $B^{\pm} \to [f]_{D}K^{\pm}$ decays with $D^{0}$ and $\bar{D}^{0}$ decays to common final states $f$ are given as follows:
\begin{eqnarray}
\Gamma(B^{-} \to [f]_{D}K^{-}) &=& [r_{B}^{2} + r_{D}^{2} + 2r_{B}r_{D}\cos(-\phi_{3}+\delta)]|A_{B}|^{2}|A_{D}|^{2} \nonumber \\
\Gamma(B^{+} \to [\bar{f}]_{D}K^{+}) &=& [r_{B}^{2} + r_{D}^{2} + 2r_{B}r_{D}\cos(\phi_{3}+\delta)]|A_{B}|^{2}|A_{D}|^{2}, \nonumber
\end{eqnarray}
where
\begin{eqnarray}
r_{B} \equiv \left|\frac{\bar{A}_{B}}{A_{B}}\right|, \:\:\:\:\:\: r_{D} \equiv \left|\frac{\bar{A}_{D}}{A_{D}}\right|, \:\:\:\:\:\: \delta \equiv \delta_{B} + \delta_{D} \nonumber
\end{eqnarray}
and $\delta_{B}$ and $\delta_{D}$ are the strong phase differences between the two $B$ and $D$ decays, respectively. The modulus of the amplitude, $|A_{B}|^{2}$ can be measured using a flavor specific $D^{0}$ decay mode. If we use a $D$ decay mode in which $A_{D}$ and $r_{D}$ are known, the above 2 equations have 3 unknowns($\phi_{3}$, $r_{B}$, $\delta$). However, using two final states $f_{1}$ and $f_{2}$, there are 4 equations and  4 unknowns($\phi_{3}$, $r_{B}$, $\delta_{1}$, $\delta_{2}$), which can be solved for $\phi_{3}$. Using multiple decay modes for $D \to f_{i}$, the value of $\phi_{3}$, and other unknowns, can be extracted from a fit. The suppressed decay $B^{-}\to[K^{+}\pi^{-}]_{D}K^{-}$ is an especially useful mode for the ADS method. The two interfering amplitudes in this decay mode are comparable, and large CP violating asymmetries can be expected. This decay mode is thus sensitive to the value of $\phi_{3}$.  

{\begin{center}
\section{\boldmath Analysis}
\end{center}}
In this paper, we report an analysis of the suppressed decay $B^{\pm}\to[K^{\mp}\pi^{\pm}]_{D}K^{\pm}$. We also analyzed the suppressed decay $B^{\pm}\to[K^{\mp}\pi^{\pm}]_{D}\pi^{\pm}$. In addition, the allowed decays $B^{\pm}\to[K^{\pm}\pi^{\mp}]_{D}K^{\pm}$ and $B^{\pm}\to[K^{\pm}\pi^{\mp}]_{D}\pi^{-}$ are used as control samples to reduce systematic uncertainties. The same selection criteria for the suppressed decay modes are applied to the control samples whenever possible. Throughout this report, charge conjugate states are implied except where explicitly mentioned and we denote the analyzed decay modes as follows.
\begin{eqnarray}
&\textrm{Suppressed decay}& B^{-}\to[K^{+}\pi^{-}]_{D}h^{-}: \:\:\:\: B^{-}\to D_{cs}h^{-} \nonumber \\
&\textrm{Allowed decay}& B^{-}\to[K^{-}\pi^{+}]_{D}h^{-}: \:\:\:\: B^{-}\to D_{f}h^{-} (h = K, \pi) \nonumber
\end{eqnarray}
The results are based on a data sample containing $274$ million $B\bar B$ pairs, collected with the Belle detector at KEKB asymmetric energy $e^{+}e^{-}$ collider operating at the $\Upsilon(4S)$ resonance. 
The Belle detector is a large-solid-angle magnetic
spectrometer that consists of a silicon vertex detector (SVD),
a 50-layer central drift chamber (CDC), an array of
aerogel threshold \v{C}erenkov counters (ACC),
a barrel-like arrangement of time-of-flight
scintillation counters (TOF), and an electromagnetic calorimeter (ECL)
comprised of CsI(Tl) crystals located inside
a super-conducting solenoid coil that provides a 1.5~T
magnetic field.  An iron flux-return located outside of
the coil is instrumented to detect $K_L^0$ mesons and to identify
muons (KLM).  The detector is described in detail elsewhere~\cite{Belle}.
Two different inner detector configurations were used. For the first sample
of 152 million $B\bar{B}$ pairs, a 2.0 cm radius beampipe
and a 3-layer silicon vertex detector were used;
for the latter 122 million $B\bar{B}$ pairs,
a 1.5 cm radius beampipe, a 4-layer silicon detector
and a small-cell inner drift chamber were used\cite{Ushiroda}.

\subsection{\boldmath Event selection}
$D$ mesons are reconstructed by combining two oppositely charged tracks. These charged tracks are required to have a point of closest approach to the beam line within $\pm5 \ {\rm mm}$ of the interaction point in the direction perpendicular to the beam axis$(dr)$ and $\pm5 \ {\rm cm}$ in the direction parallel to the beam axis$(dz)$. A $K/\pi$ likelihood ratio $P(K/\pi) = \mathcal{L}_{K}/(\mathcal{L}_{K} + \mathcal{L}_{\pi})$ is formed for each track, where $\mathcal{L}_{K}$ and $\mathcal{L}_{\pi}$ are kaon and pion likelihoods. We used the particle identification requirement  $P(K/\pi) > 0.4$ and $P(K/\pi) < 0.7$ for kaons and pions from $D \to K\pi$ decays, respectively. $D$ candidates are required to have an invariant mass within $\pm2.5\sigma$ of the nominal $D^{0}$ mass: 1.850 GeV/$c^{2}$ $< M(K\pi) <$ 1.879 GeV/$c^{2}$. To improve the momentum determination, tracks from the $D$ candidate are refitted according to the nominal $D^{0}$ mass hypothesis and the reconstructed vertex position (a mass-and-vertex-constrained fit).

$B$ mesons are reconstructed by combining $D$ candidates with primary charged hadron candidates. For the charged tracks, we require $P(K/\pi) > 0.6$ for the kaon in $B^{-} \to DK^{-}$ and $P(K/\pi) < 0.2$ for the pion in $B^{-} \to D\pi^{-}$. The signal is identified by two kinematic variables, the energy difference $\Delta E = E_{D} + E_{K^{-}(\pi^{-})} - E_{\rm beam}$ and the beam-energy-constrained mass $M_{\rm bc} = \sqrt{E^{2}_{\rm beam} - (\vec{p}_{D} + \vec{p}_{K^{-}(\pi^{-})})^{2}}$, where $E_{D}$ is the energy of the $D$ candidate, $E_{K^{-}(\pi^{-})}$ is the energy of the $K^{-}(\pi^{-})$ and $E_{\rm beam}$ is the beam energy, in the cm frame. $\vec{p}_{D}$ and $\vec{p}_{K^{-}(\pi^{-})}$ are the momenta of the $D$ and $K^{-}(\pi^{-})$ in the cm frame. We define the signal region as 5.27 GeV/$c^{2}$ $< M_{\rm bc} <$ 5.29 GeV/$c^{2}$ and -0.04 GeV $< \Delta E <$ 0.04 GeV. In the case of multiple candidates per event, we choose the best candidate on the basis of a $\chi^{2}$ determined from the difference between the measured and nominal values of $M_{D}$ and $M_{\rm bc}$.  
  
\subsection{\boldmath $q\bar{q}$ continuum suppression}
To suppressed the large background from the two-jet like $e^{+}e^{-} \to q\bar{q}(q = u, d, s, c)$ continuum processes, variables that characterize the event topology are used. We construct a Fisher discriminant of Fox-Wolfram moments called the Super-Fox-Wolfram$(SFW)$~\cite{fisher}\cite{sfw}, 
where the Fisher coefficients are optimized by maximizing the separation between $B\bar{B}$ events and continuum events. Furthermore, $\cos\theta_{B}$, the angle in the cm system between the $B$ flight direction with respect to the beam axis is used as another variable to distinguish $B\bar{B}$ events from continuum events. These two independent variables, $SFW$ and $\cos\theta_{B}$ are combined to form a likelihood ratio($LR$),
\begin{eqnarray}
LR &=& {\cal L}_{\rm sig}/({\cal L}_{\rm sig} + {\cal L}_{\rm cont}) \nonumber \\
 {\cal L}_{\rm sig(cont)} &=& {\cal L}_{\rm sig(cont)}^{SFW} \times {\cal L}_{\rm sig(cont)}^{\cos\theta_{B}}, \nonumber
\end{eqnarray}
where ${\cal L}_{\rm sig}$ and ${\cal L}_{\rm cont}$ are likelihoods defined from $SFW$ and $\cos\theta_{B}$ distributions for signal and continuum backgrounds, respectively. We optimized the $LR$ requirement by maximizing a figure of merit, $S/\sqrt{S +N}$, where $S$ and $N$ denote the expected number of signal and background in the signal region. For $B^{-} \to D_{cs}K^{-}(\pi^{-})$ we require $LR > 0.85(> 0.75)$, which retains $44.8\%(57.6\%)$ signal events and removes $96.2\%(93.2\%)$ of the continuum background. 

\subsection{\boldmath Peaking backgrounds}
\begin{figure}[p]
\begin{center}
\includegraphics[width=0.45\textwidth]{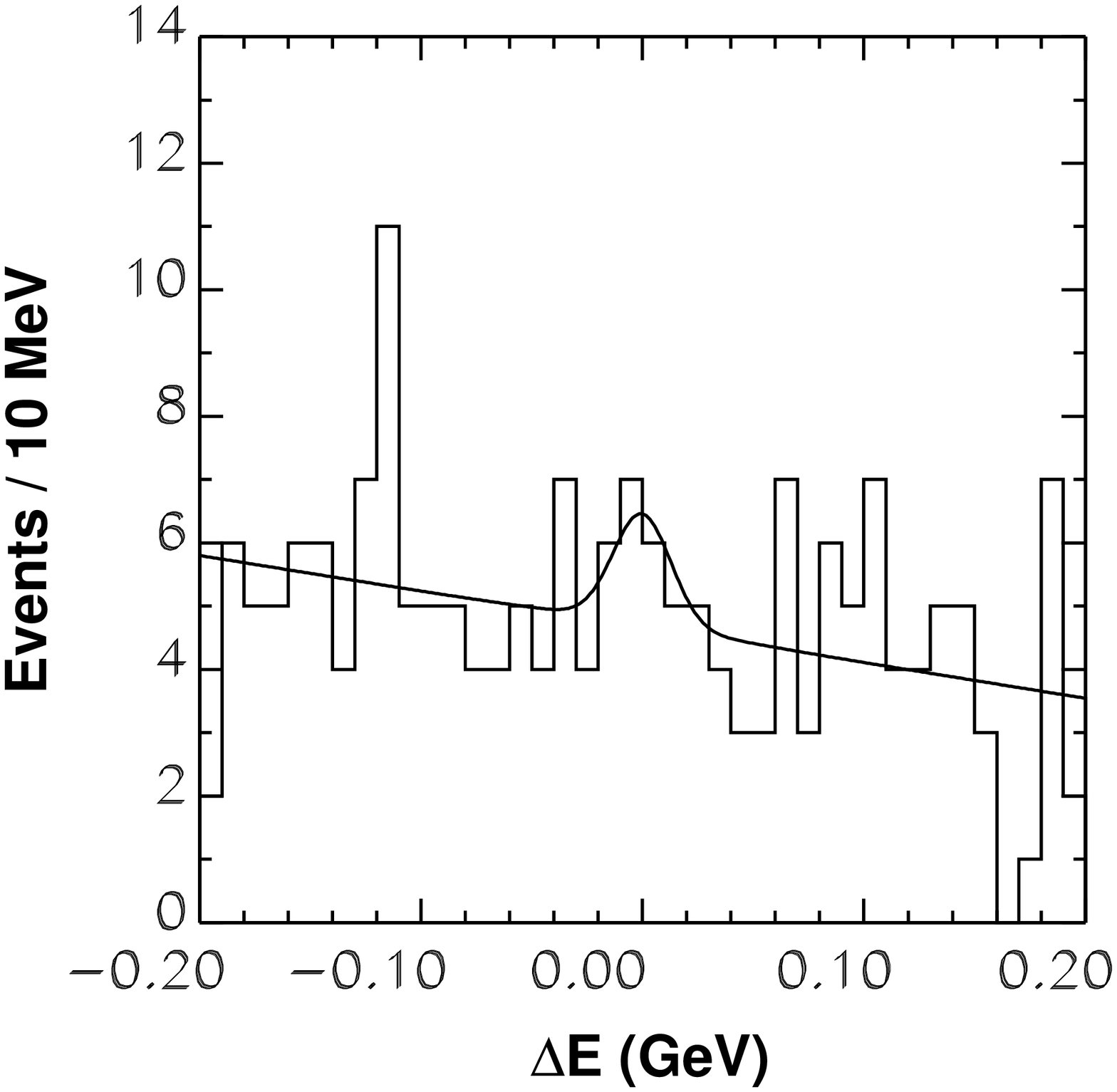}
\includegraphics[width=0.45\textwidth]{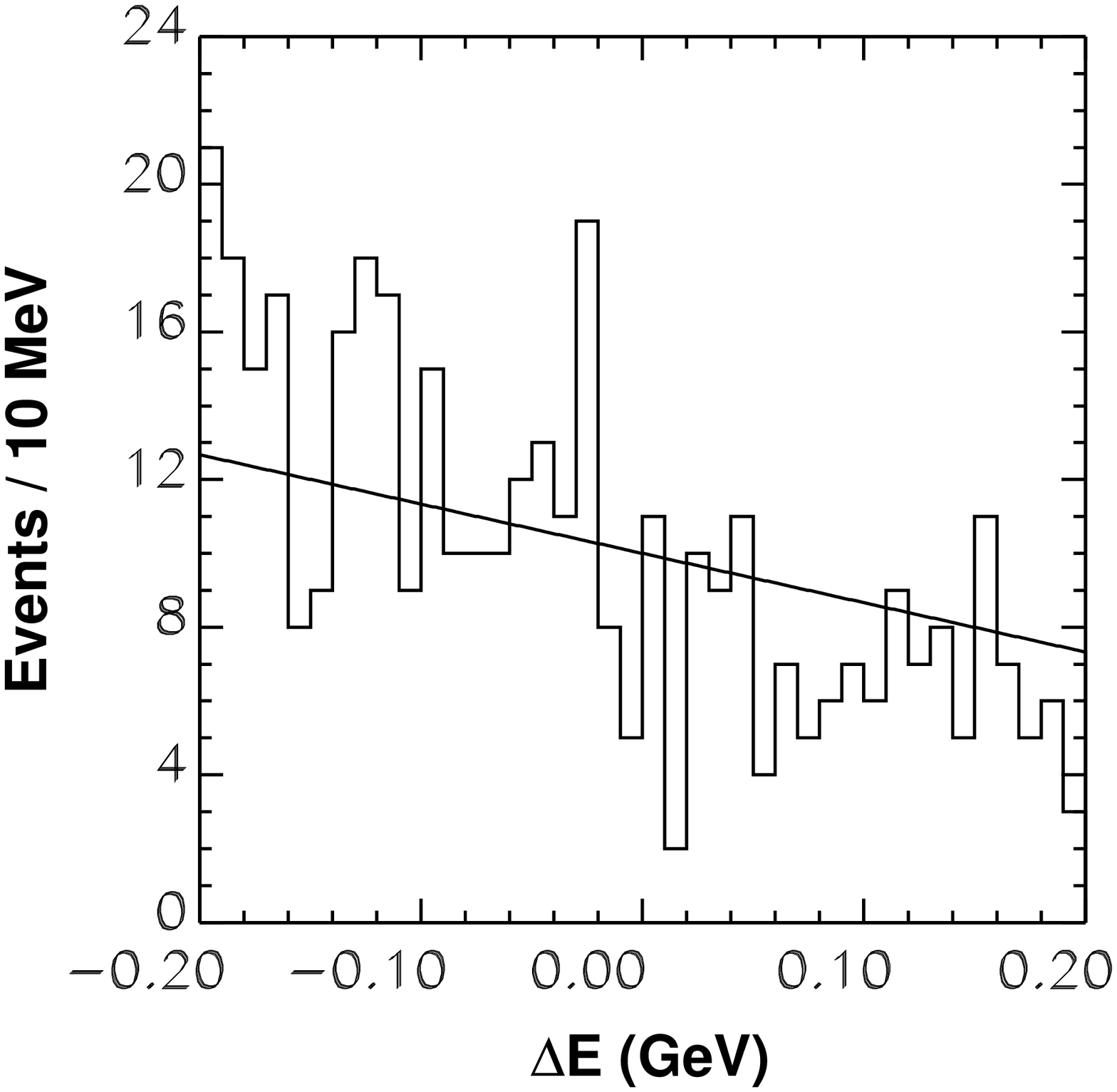}
\caption{$\Delta E$ distributions for events in the $D^{0}$ mass sideband for $B^{-}\to D_{cs}K^{-}$(left) and $B^{-}\to D_{cs}\pi^{-}$(right). The signal shapes are modeled using the results of the $B^{-}\to D_{f}h^{-}(h=K,\pi)$ fit.}
\label{fig:peakbkg}
\end{center}
\end{figure}

For $B^{-} \to D_{cs}K^{-}$, one can have a contribution from $B^{-}\to D^{0}\pi^{-}$, $D^{0} \to K^{+}K^{-}$, which has the same final state and can peak under the signal. In order to reject these events, we veto events that satisfy 1.843 GeV/$c^{2} < M(KK) <$ 1.894 GeV/$c^{2}$. The allowed decay $B^{-}\to D_{f}h^{-}$ can also be a peaking background for the suppressed decay modes due to $K\pi$ misidentification. Therefore, we veto events for which the invariant mass of the $K\pi$ pair is inside the $D$ mass cut window when the mass assignments are exchanged. Furthermore, three-body charmless decays $B^{-} \to K^{+}K^{-}\pi^{-}$ and $B^{-} \to K^{+}\pi^{-}\pi^{-}$ can peak inside the signal region for $B^{-} \to D_{cs}K^{-}$ and $B^{-} \to D_{cs}\pi^{-}$, respectively. These peaking backgrounds are estimated from the $\Delta E$ distributions of events in a $D$ mass sideband, defined as 1.808 GeV/$c^{2}$ $< M(K\pi) <$ 1.836 GeV/$c^{2}$ and 1.893 GeV/$c^{2}$ $< M(K\pi) <$ 1.922 GeV/$c^{2}$, which are shown in Fig.\ref{fig:peakbkg}.  For $B^{-} \to D_{cs}\pi^{-}$, the peaking background estimated by fitting the plot is consistent with zero. Since the Standard Model prediction for the $B^{-} \to K^{+}\pi^{-}\pi^{-}$ branching fraction is smaller than $10^{-11}$\cite{sm}, this background contribution is ignored. On the other hand, for $B^{-} \to D_{cs}K^{-}$, the estimated peaking background is $3.1 \pm 2.9$ events inside the $\Delta E$ signal region after scaling to the $D$ mass signal region. As a check, we naively estimate the expected background from the measured $B^{-} \to K^{+}K^{-}\pi^{-}$ mode. According to \cite{hhh}, the $B^{-} \to K^{+}K^{-}\pi^{-}$ yield is $94 \pm 23$ events with an efficiency of $13.8\%$ ($78.7 \ {\rm fb}^{-1}$). Using this result, the estimated background is
\begin{eqnarray}
94 \times \frac{257.1\:{\rm fb}^{-1}}{78.7\:{ \rm fb}^{-1}} \times \frac{\textrm{area}_{D}}{\textrm{area}_{Dalitz}} \times \frac{\textrm{eff}_{DK}}{\textrm{eff}_{KK\pi}} \sim 2.9 \textrm{ events},\nonumber
\end{eqnarray} 
where we assumed that the $B^{-} \to K^{+}K^{-}\pi^{-}$ yield is uniformly distributed over the Dalitz plot, and $\textrm{area}_{D}/\textrm{area}_{Dalitz}$ is the ratio between the $D$ mass cut area and the Dalitz plot area, and $\textrm{eff}_{DK}/\textrm{eff}_{KK\pi}(=17.5/13.8)$ is the ratio of the $B^{-} \to D_{cs}K^{-}$ efficiency to the $B^{-} \to K^{+}K^{-}\pi^{-}$ efficiency. This naive estimate is consistent with the estimate from the $D^{0}$ mass sideband. Therefore, we subtract $3.1 \pm 2.9$ events from the observed $B^{-}\to D_{cs}K^{-}$ yield.

After applying all the cuts, the signal efficiencies are $17.5\%$ and $24.6\%$ for $B^{-}\to D_{cs}K^{-}$ and $B^{-}\to D_{cs}\pi^{-}$, respectively. The signal yields are extracted by fitting the $\Delta E$ distributions.

\subsection{\boldmath Fitting the $\Delta E$ distributions} 
Backgrounds from decays such as $B^{-} \to D\rho^{-}$ and $B^{-} \to D^{*}\pi^{-}$ are distributed in the negative $\Delta E$ region and make a small contribution to the signal region. The shape of this $B\bar{B}$ background is modeled as a smoothed histogram from generic Monte Carlo (MC) samples. The continuum background populates the entire $\Delta E$ region. The shape of the  continuum background is modeled as a linear function. The slope is determined from the $\Delta E$ distribution of the $M_{bc}$ sideband data (5.20 GeV/$c^{2} < M_{bc} <$ 5.26 GeV/$c^{2}$).

The $\Delta E$ fitting function is the sum of two Gaussians for the signal, the linear function for the continuum, and the smoothed histogram for the $B\bar{B}$ background distribution. 

\begin{figure}[p]
\begin{center}
\includegraphics[width=0.47\textwidth]{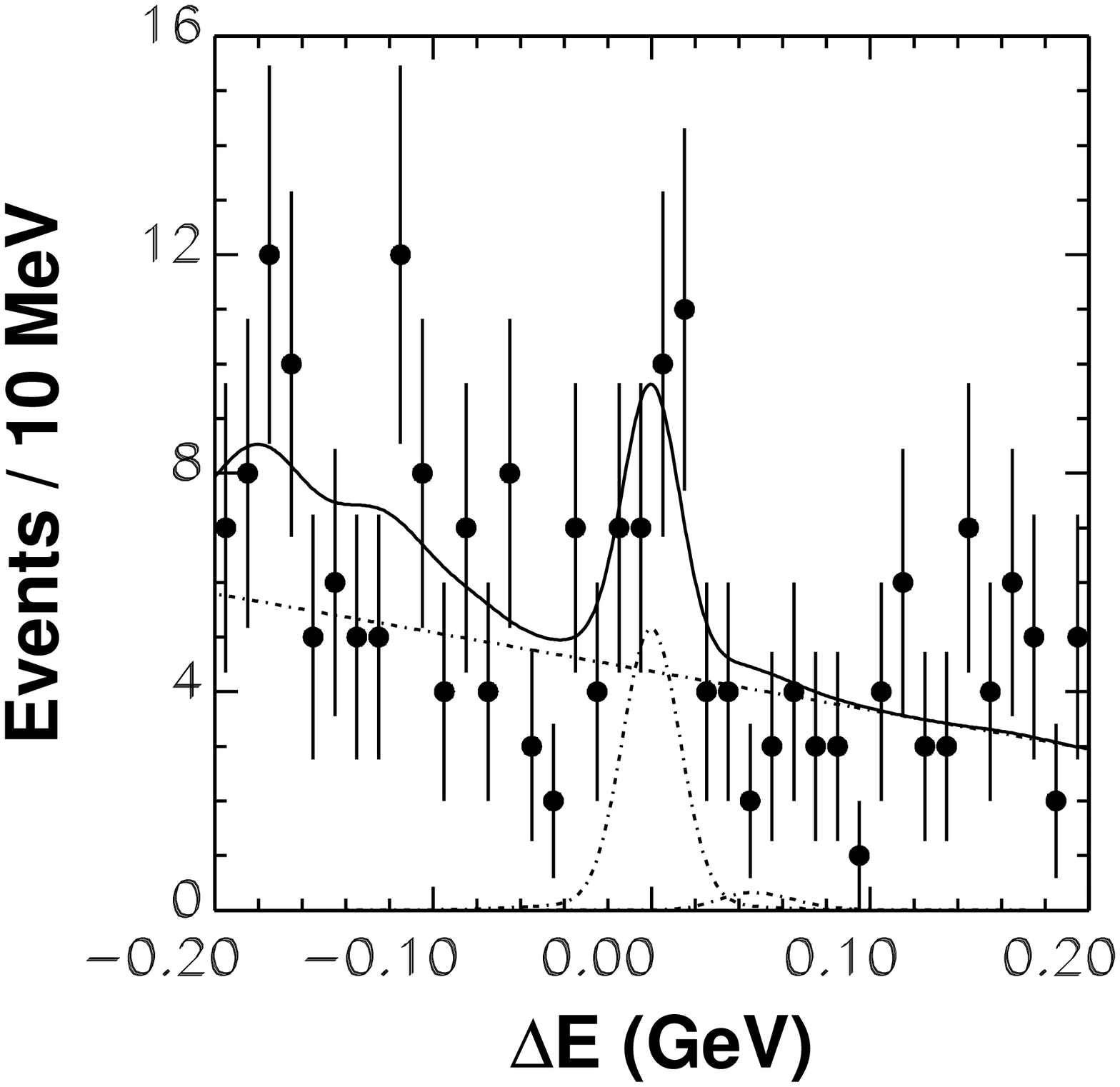}
\includegraphics[width=0.47\textwidth]{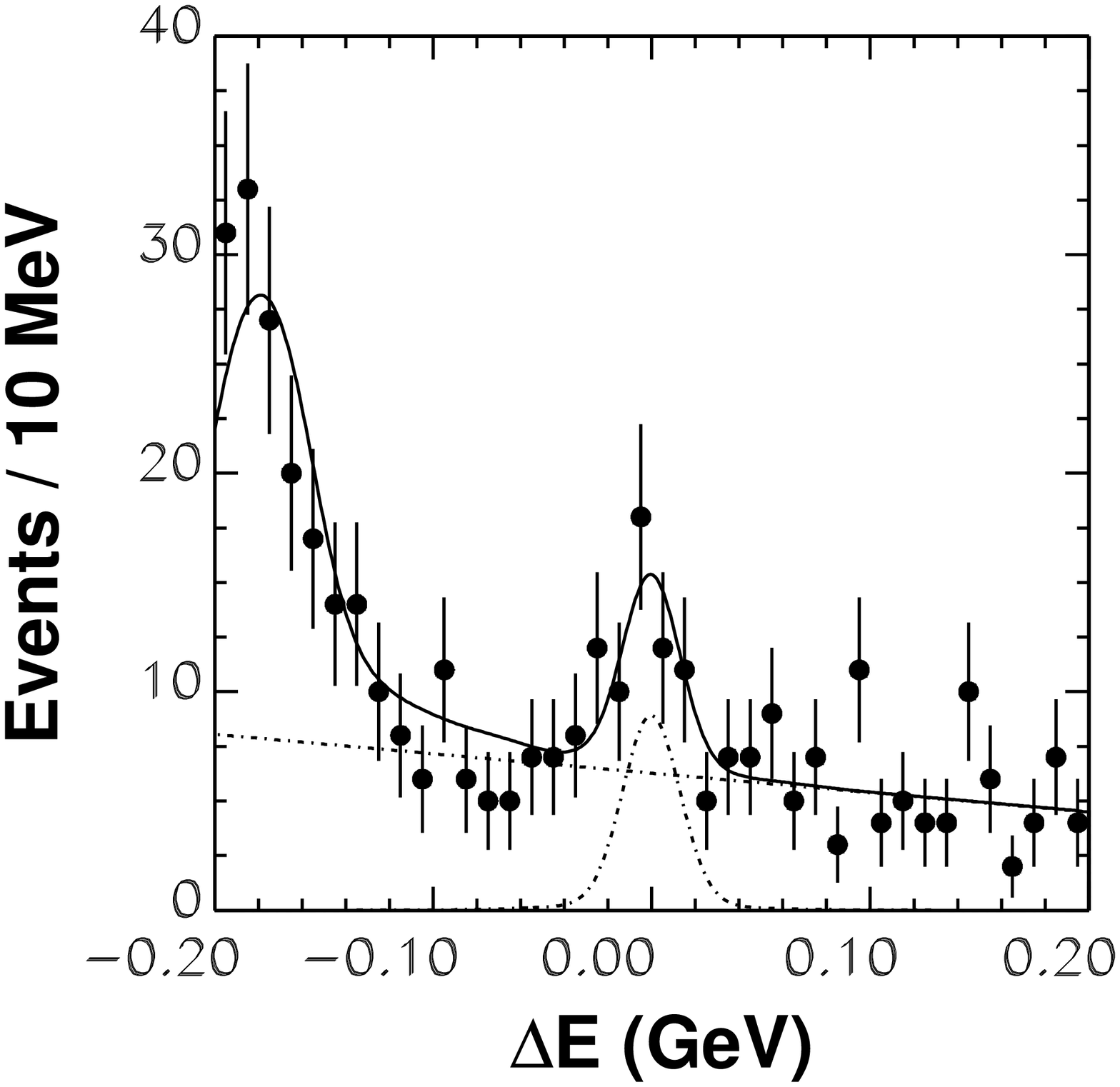}
\includegraphics[width=0.47\textwidth]{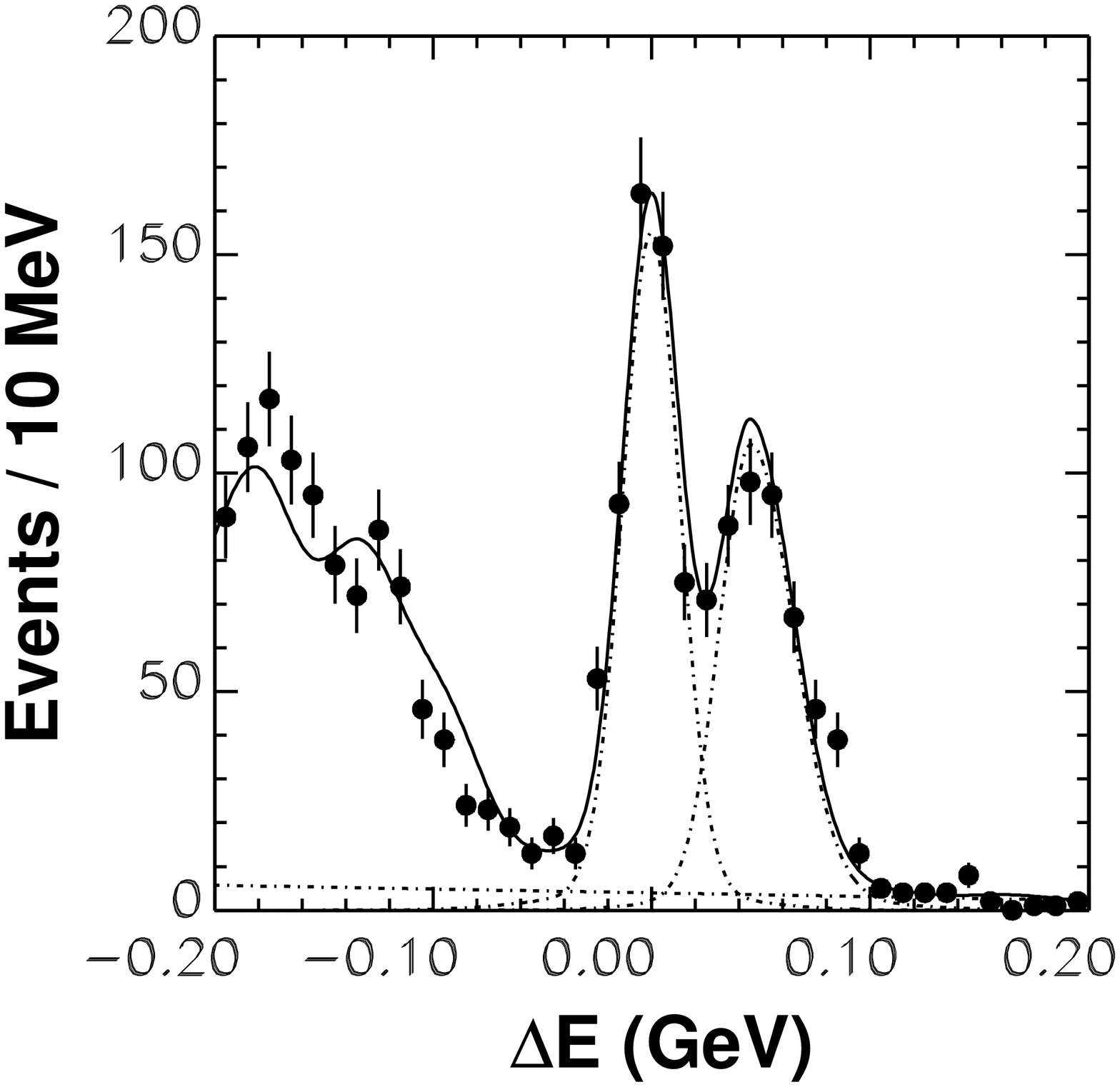}
\includegraphics[width=0.47\textwidth]{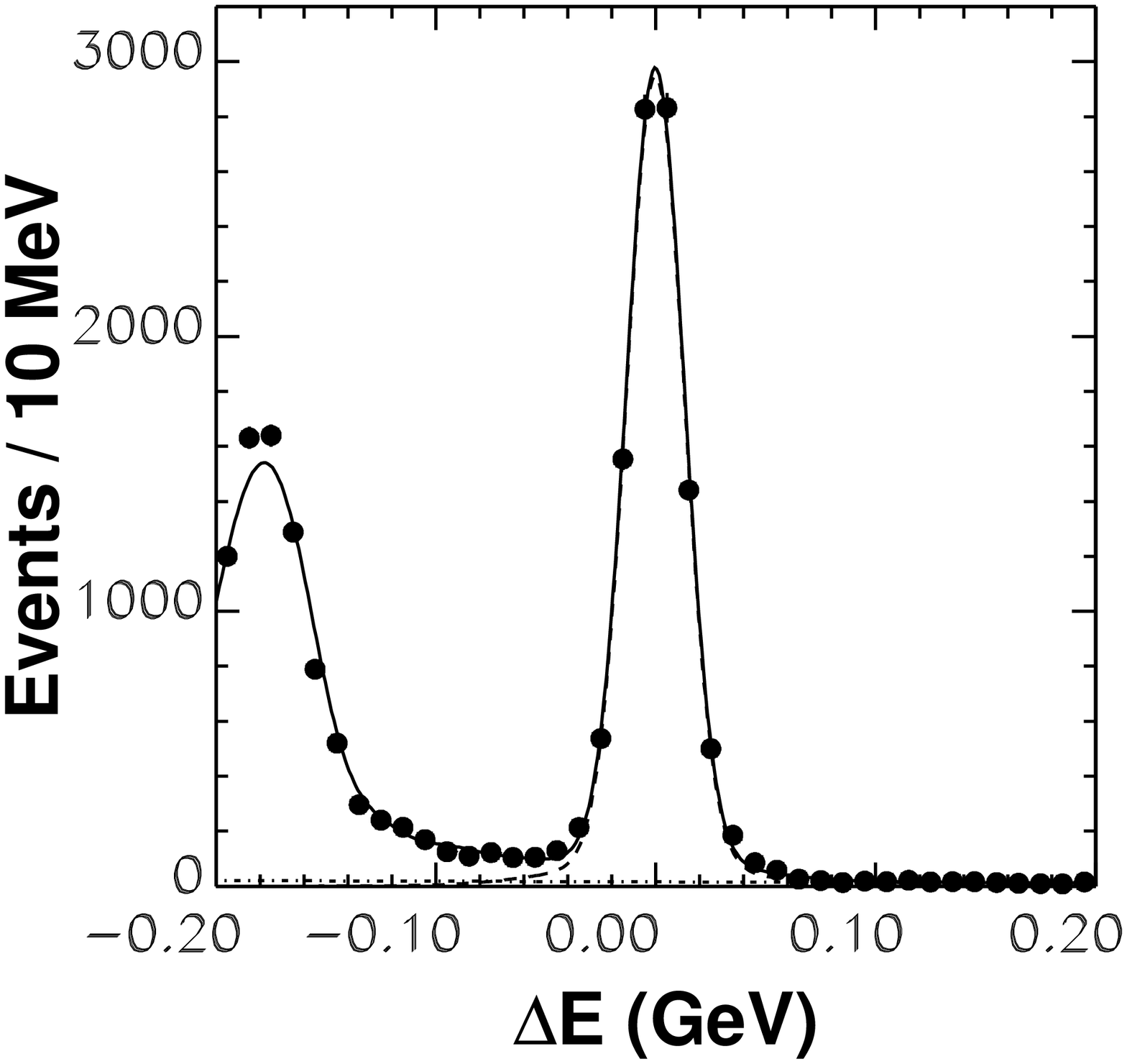}
\caption{$\Delta E$ fit results for $B^{-}\to D_{cs}K^{-}$(top-left), $B^{-}\to D_{cs}\pi^{-}$(top-right), $B^{-}\to D_{f}K^{-}$(bottom-left), and $B^{-}\to D_{f}\pi^{-}$(bottom-right). The charge conjugate modes are included for these plots.}
\label{fig:fitting}
\end{center}
\end{figure}

In the fit to the $\Delta E$ distribution of $B^{-} \to D_{f}\pi^{-}$, the free parameters are the position, width and area of the signal peak, and the normalizations of continuum and $B\bar{B}$ backgrounds. The ratio of the two Gaussians of the signal is fixed from the signal MC. 
For the $B^{-} \to D_{f}K^{-}$ fit, the position and width of the signal peak are fixed from the $B^{-} \to D_{f}\pi^{-}$ fit results. To fit the feed-across from $D_{f}\pi^{-}$, we use a Gaussian shape where the left and right sides of the peak have different widths since the shift caused by wrong mass assignment makes the shape asymmetric. The shape parameters of this function are fixed at values determined by the fit to the $B^{-} \to D_{f}\pi^{-}$ distribution using a kaon mass hypothesis for the prompt pion. The areas of signal and feed-across from $D\pi^{-}$, and the normalizations of continuum and $B\bar{B}$ backgrounds are floated in the fit. 
For $B^{-} \to D_{cs}K^{-}$ and $B^{-} \to D_{cs}\pi^{-}$, the signal and $B\bar{B}$ background shapes are modeled using the fit results of the $B^{-} \to D_{f}K^{-}$ and $B^{-} \to D_{f}\pi^{-}$ modes, respectively.  The area of the feed-across from $D_{cs}\pi^{-}$ is estimated as the measured yield of $B^{-} \to D_{cs}\pi^{-}$ multiplied by the $\pi$ to $K$ misidentification probability. However, the areas of the signal and the normalizations of continuum and $B\bar{B}$ backgrounds are floated.  The fit results are shown in Fig.\ref{fig:fitting}. The numbers of events for $B^{-} \to D_{cs}h^{-}$ and $D_{f}h^{-}$, and the statistical significances of the $B^{-} \to D_{cs}h^{-}$ signals are given in Table \ref{tab:yield}. The statistical significance is defined as $\sqrt{-2\ln({\cal L}_{0}/{\cal L}_{\rm max})}$, where ${\cal L}_{\rm max}$ is the maximum likelihood in the $\Delta E$ fit and ${\cal L}_{0}$ is the likelihood when the signal yield is constrained to be zero. The uncertainty in the peaking background contribution is taken into account in the statistical significance calculation. The statistical significance of the $B^{-} \to D_{cs}\pi^{-}$ signal is over $5.0\sigma$.

\begin{table}[h]
\caption{Signal yields and efficiency. For the $B^{-}\to D_{cs}K^{-}$ signal yield, the peaking background contribution has been subtracted.}
\begin{tabular}{ l | c c c c}
\hline
\hline
Mode & Product branching\hspace{2mm} & \hspace{2mm}Efficiency\hspace{2mm} & \hspace{2mm}Signal Yield\hspace{2mm} & Statistical\\
& fraction from PDG & ($\%$) & & significance \\
\hline
$B^{-}\to D_{cs}K^{-}$ & $-$ & $17.5 \pm 0.2$ & $14.7^{+8.0}_{-7.3}$ & 2.7 \\
$B^{-}\to D_{cs}\pi^{-}$ & $(6.9\pm0.7) \times 10^{-7}$ & $24.6 \pm 0.2$ & $30.7^{+9.1}_{-8.4}$ & 5.8 \\
$B^{-}\to D_{f}K^{-}$ & $(1.4 \pm 0.2) \times 10^{-5}$ & $17.5 \pm 0.3$ & $535.0^{+18.8}_{-18.2}$  \\
$B^{-}\to D_{f}\pi^{-}$ & $(1.9 \pm 0.1) \times 10^{-4}$ & $24.7 \pm 0.2$ & $10178^{+105}_{-104}$  \\
\hline
\hline
\end{tabular}
\label{tab:yield}
\end{table}

{\begin{center}
\section{\boldmath Results}
\end{center}}
\subsection{\boldmath Branching fraction of suppressed decay modes}

The branching fractions for $B^{-}\to D_{cs}h^{-}(h=K,\pi)$ are determined as
\begin{eqnarray}
{\cal B}(B^{-}\to D_{cs}h^{-}) = {\cal B}(B^{-}\to D_{f}h^{-}) \times \frac{N_{D_{cs}h}}{N_{D_{f}h}}, \nonumber
\end{eqnarray}
where $N_{D_{cs}h}$ and $N_{D_{f}h}$ are the number of $B^{-}\to D_{cs}h^{-}$ signal events and $B^{-}\to D_{f}h^{-}$ signal events. The product branching fractions for $B^{-}\to D_{f}h^{-}$, calculated from the world averages for the branching fractions \cite{pdg}, are given in Table \ref{tab:yield}. Using these, the branching fractions for the suppressed decays $B^{-}\to D_{cs}h^{-}$ are found to be
\begin{eqnarray}
&{\cal B}(B^{-}\to[K^{+}\pi^{-}]_{D}K^{-})& =  (3.9^{+2.1}_{-1.9}(stat) \pm0.2(sys) \pm0.6(PDG)) \times 10^{-7}, \nonumber \\
&{\cal B}(B^{-}\to[K^{+}\pi^{-}]_{D}\pi^{-})& = (5.7^{+1.7}_{-1.6}(stat) \pm0.3(sys) \pm0.3(PDG)) \times 10^{-7}. \nonumber
\end{eqnarray}
Most of the systematic uncertainties from the detection efficiencies and the particle identification cancel when taking the ratios, since the kinematics of the $B^{-}\to D_{cs}h^{-}$ and $B^{-}\to D_{f}h^{-}$ processes are similar. The systematic errors are due to the uncertainty in the yield extraction and the efficiency difference between $B^{-}\to D_{cs}h^{-}$ and $B^{-}\to D_{f}h^{-}$. The uncertainties in the signal shapes and the $q\bar q$ background shapes are determined by varying the shape of the fitting function by $\pm1\sigma$. The uncertainties in the $B\bar B$ background shapes are determined by fitting the $\Delta E$ distribution in the region -0.07 GeV $< \Delta E <$ 0.20 GeV ignoring the $B\bar B$ background contributions. The uncertainties in the efficiency differences are determined by the signal MC. The total systematic errors are obtained as the quadratic sum of those uncertainties. The results are shown in Table \ref{tab:sys}.

The uncertainties in the branching fractions are statistics-dominated. For the $B^{-}\to D_{cs}K^{-}$ branching fraction, we set an upper limit at the $90\%$ confidence level as
\begin{eqnarray}
&{\cal B}(B^{-}\to D_{cs}K^{-})& < 7.6 \times 10^{-7} (90\% \:{\rm C.L.}), \nonumber
\end{eqnarray}
where we took the likelihood function as a single gaussian with width given by the quadratic sum of the statistical and systematic errors, and the area is normalized in the physical region of positive branching fraction. 

\begin{table}[h]
\caption{Systematic uncertainties for $B^{-}\to DK^{-}$ and $B^{-}\to D\pi^{-}$.}
\begin{tabular}{ l | c c | c c }
\hline
\hline
& \multicolumn{4}{c}{Systematic error($\%$)} \\
\hline
Source & $D_{cs}K^{-}$ &  $D_{f}K^{-}$ &  $D_{cs}\pi^{-}$ &  $D_{f}\pi^{-}$ \\
\hline
$B\bar B$ background shape & $\pm2.1$ & $\pm1.0$ & $\pm4.6$ & $\pm1.6$ \\
$q\bar q$ background shape & $\pm3.6$ & $\pm0.4$ & $\pm1.9$ & $\pm0.1$ \\
Signal shape & $\pm0.6$ & $\pm0.4$ & $\pm1.4$ & $\pm0.2$ \\
Feed-across shape& $\pm1.4$ & $\pm1.0$ & $-$ & $-$ \\ 
Efficiency difference  & \multicolumn{2}{c|}{$\pm1.5$} & \multicolumn{2}{c}{$\pm1.3$} \\
PDG Normalization & \multicolumn{2}{c|}{$\pm14.3$} & \multicolumn{2}{c}{$\pm 5.3$} \\
\hline
Total & \multicolumn{2}{c|}{$\pm4.9 \pm14.3_{(PDG)}$} & \multicolumn{2}{c}{$\pm5.5\pm5.3_{(PDG)}$} \\
\hline
\hline
\end{tabular}
\label{tab:sys}
\end{table}

\subsection{\boldmath Ratio of branching fractions $R_{Dh}$}
We define the ratio
\begin{eqnarray}
R_{Dh} &\equiv& \frac{{\cal B}(B^{-}\to D_{cs}h^{-}) + {\cal B}(B^{+}\to D_{cs}h^{+})}{{\cal B}(B^{-}\to D_{f}h^{-}) + {\cal B}(B^{+}\to D_{f}h^{+})}  \:\:\:\:\:\:\:(h = K,\pi)\nonumber \\
\nonumber \\
& = & \frac{N_{D_{cs}h}}{N_{D_{f}h}}. \nonumber 
\end{eqnarray}
The ratios $R_{Dh}$ are determined as follows
\begin{eqnarray}
R_{DK} &=& (2.8^{+1.5}_{-1.4}(stat) \pm 0.1(sys)) \times 10^{-2}, \nonumber \\
R_{D\pi} &=& (3.0^{+0.9}_{-0.8}(stat) \pm 0.2(sys)) \times 10^{-3} \nonumber
\end{eqnarray}
and
\begin{eqnarray}
R_{DK} < 4.7 \times 10^{-2}(90\% \:{\rm C.L.}). \nonumber
\end{eqnarray}
The ratio $R_{DK}$ is related to $\phi_{3}$ by
\begin{eqnarray}
R_{DK} = r_{B}^{2} + r_{D}^{2} + 2r_{B}r_{D}\cos\phi_{3} \cos\delta, \nonumber
\end{eqnarray}
where
\begin{eqnarray}
r_{D} = \left|\frac{A(D^{0} \to K^{+}\pi^{-})}{A(D^{0} \to K^{-}\pi^{+})}\right| = 0.060 \pm 0.003. \nonumber
\end{eqnarray}
Using the above result, we obtain a limit on $r_B$. The least restrictive limit is obtained allowing $\pm 1\sigma$ variation on $r_{D}$~\cite{pdg} and assuming maximal interference($\phi_{3} = 0^{\circ}, \delta = 180^{\circ}$ or $\phi_{3} = 180^{\circ}, \delta = 0^{\circ}$) and is found to be
\begin{eqnarray}
r_{B} &<& 0.28. \nonumber
\end{eqnarray}

\subsection{\boldmath $CP$ asymmetry}
\begin{figure}[p]
\begin{center}
\includegraphics[width=0.47\textwidth]{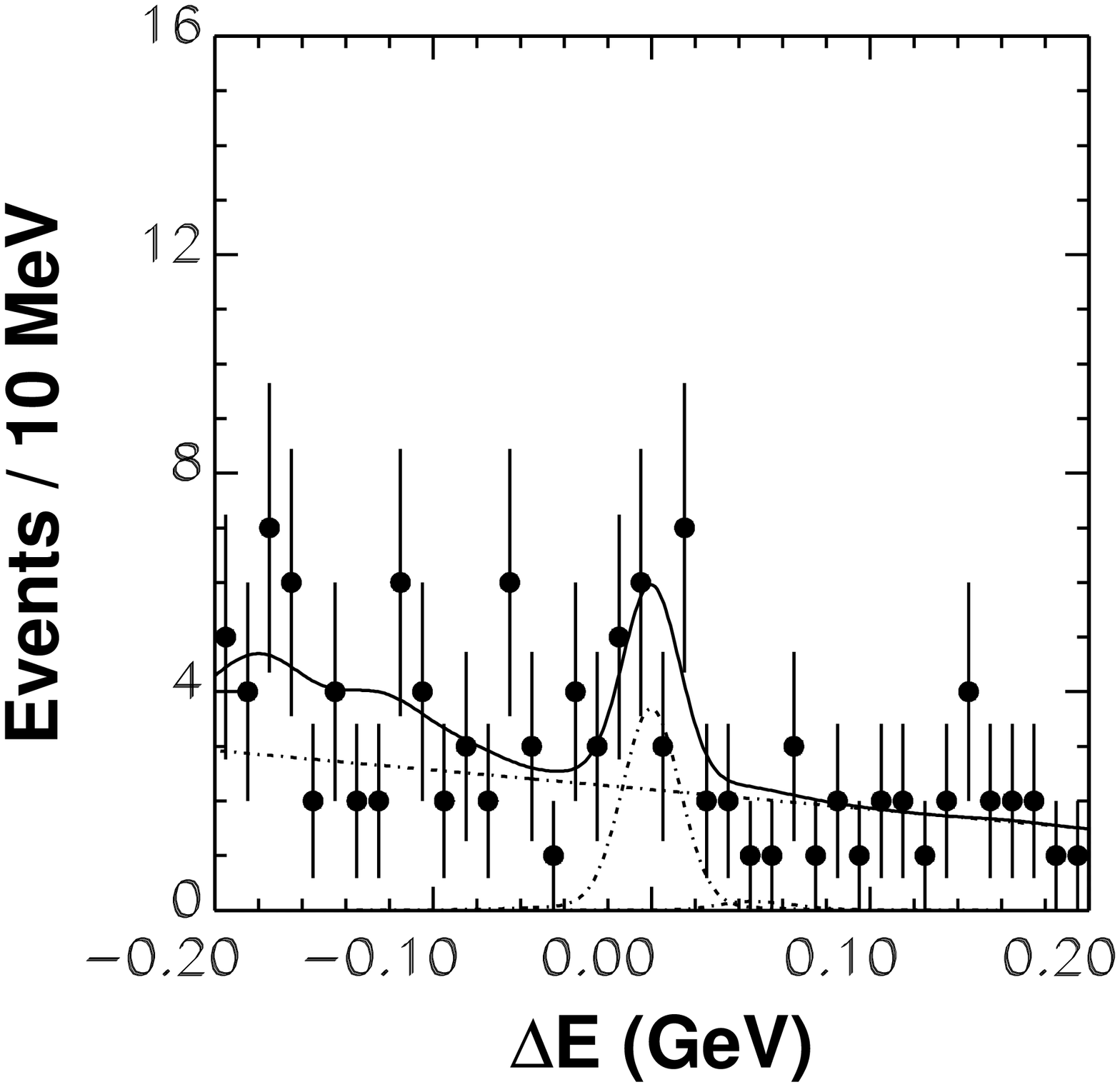}
\includegraphics[width=0.47\textwidth]{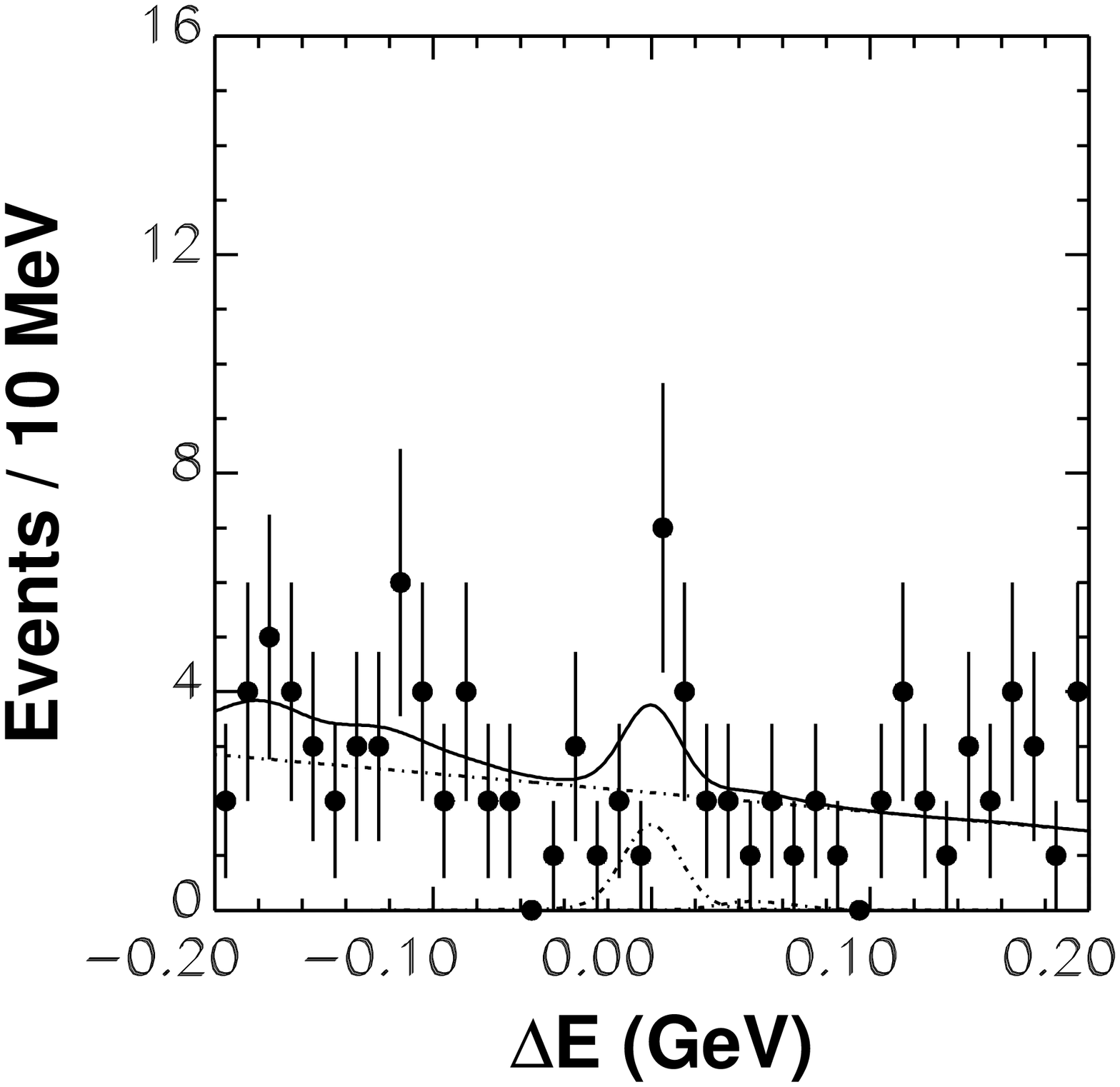}
\includegraphics[width=0.47\textwidth]{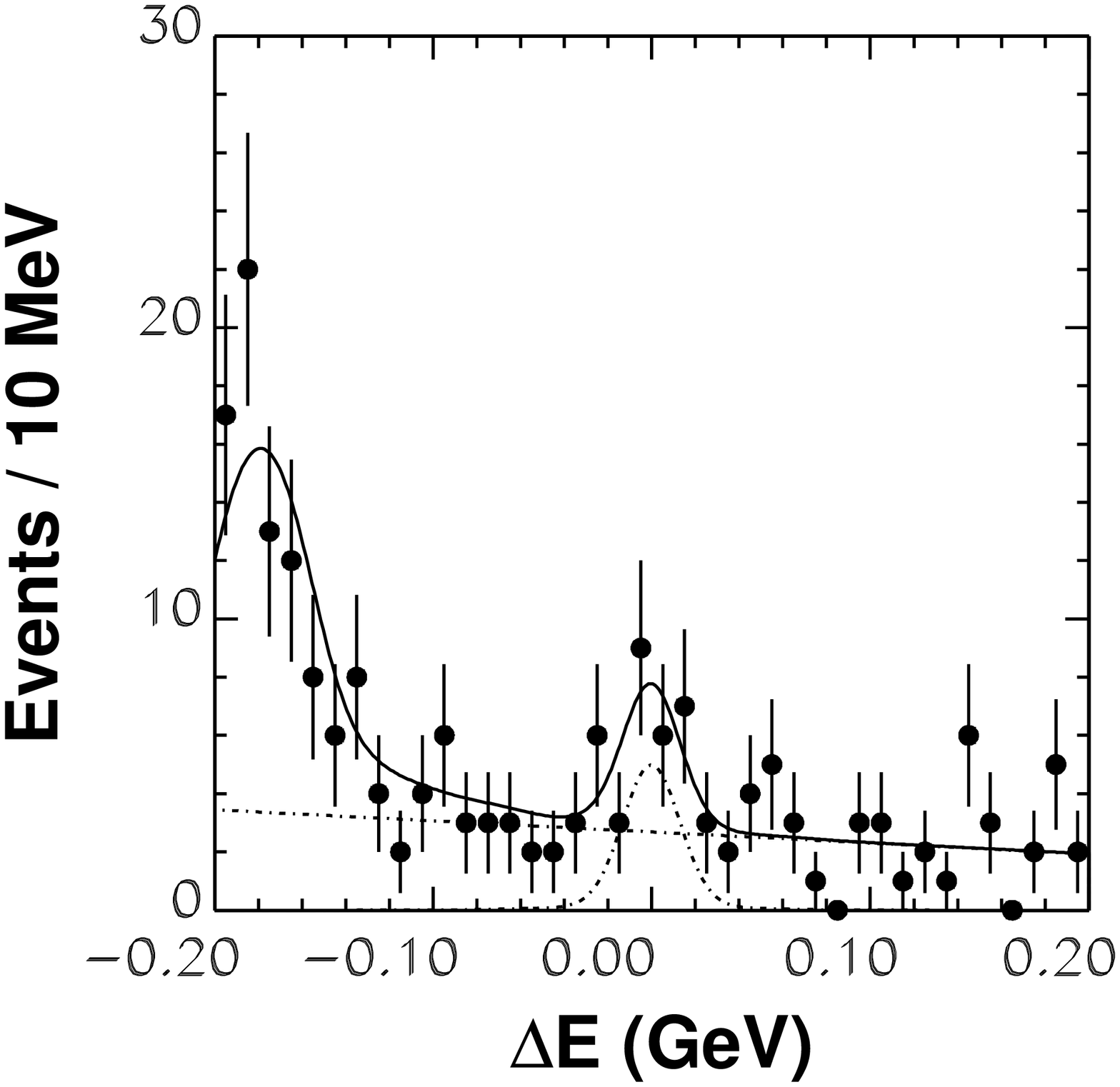}
\includegraphics[width=0.47\textwidth]{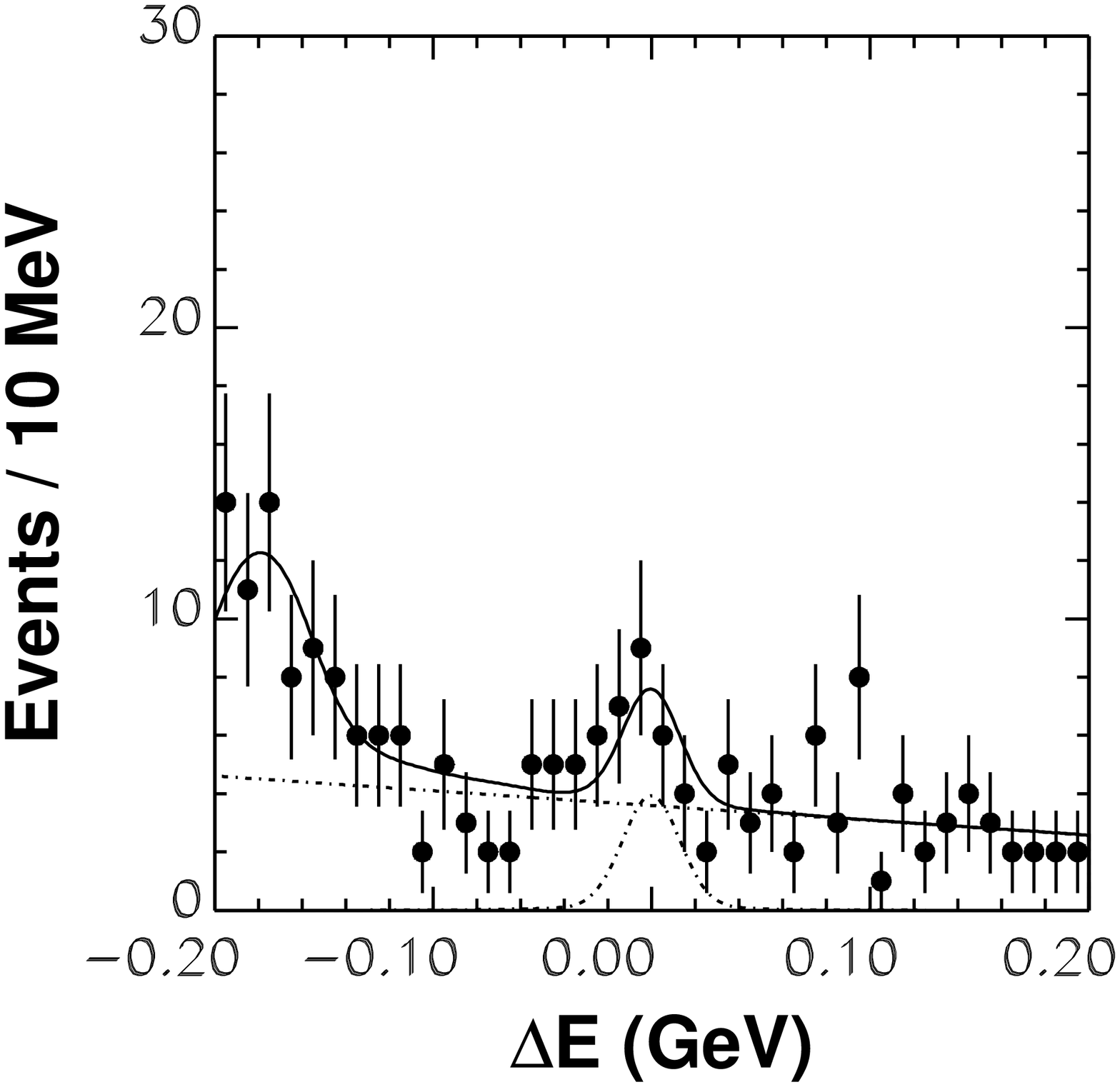}
\caption{$\Delta E$ fit results for $B^{-}\to D_{cs}K^{-}$(top-left), $B^{+}\to D_{cs}K^{+}$(top-right), $B^{-}\to D_{cs}\pi^{-}$(bottom-left), and $B^{+}\to D_{cs}\pi^{+}$(bottom-right).}
\label{fig:acp}
\end{center}
\end{figure}

We search for partial rate asymmetries ${\cal A}_{Dh}$ in $B^{\pm}\to D_{cs}h^{\pm}$ decay, fitting the $B^{+}$ and $B^{-}$ yields separately for each mode, where ${\cal A}_{Dh}$ is determined as
\begin{eqnarray}
{\cal A}_{Dh} &\equiv& \frac{{\cal B}(B^{-}\to D_{cs}h^{-}) - {\cal B}(B^{+}\to D_{cs}h^{+})}{{\cal B}(B^{-}\to D_{cs}h^{-}) + {\cal B}(B^{+}\to D_{cs}h^{+})} \:\:\:\:\:\:\:(h = K,\pi).\nonumber 
\end{eqnarray}
The peaking background for $B^{-}\to D_{cs}K^{-}$ is subtracted assuming no $CP$ asymmetry. The fit results are shown in Fig.\ref{fig:acp} and Table \ref{tab:acp}. We find
\begin{eqnarray}
{\cal A}_{DK} &=& 0.49^{+0.53}_{-0.46}(stat) \pm0.06(sys),\nonumber \\
{\cal A}_{D\pi} &=& 0.12^{+0.30}_{-0.27}(stat) \pm0.06(sys), \nonumber 
\end{eqnarray}
where the systematic uncertainty is from the intrinsic detector charge asymmetry, the $B^{+}$ and $B^{-}$ yield extraction, and the asymmetry in particle identification efficiency of prompt kaons.  The intrinsic detector charge asymmetry is determined from the $B^{\pm} \to D_{f}\pi^{\pm}$ samples. The systematic uncertainty from yield extraction is determined by varying the fitting parameters by $\pm1\sigma$.  The systematic uncertainty due to particle identification efficiency of prompt kaons is explained in \cite{dcpk}. The total systematic errors are combined as the quadratic sum of those uncertainties (Table \ref{tab:acpsys}). The measured partial rate asymmetries ${\cal A}_{Dh}$ are consistent with zero. 

\begin{table}[h]
\caption{Signal yields and partial rate asymmetries.}
\begin{tabular}{ l | c c c c c}
\hline
\hline
Mode & $N(B^{-})$ & & $N(B^{+})$ & & ${\cal A}_{Dh}$\\
\hline
$B\to D_{cs}K$ & $11.2^{+6.1}_{-5.4}$ & & $3.9^{+4.9}_{-4.3}$ & & $0.49^{+0.53}_{-0.46}\pm0.06$\\
$B\to D_{cs}\pi$ & $17.2^{+6.5}_{-5.8}$ & & $13.6^{+6.6}_{-5.9}$ & & $0.12^{+0.30}_{-0.27}\pm0.06$\\
\hline
\hline
\end{tabular}
\label{tab:acp}
\end{table}

\begin{table}[h]
\caption{Source of systematic uncertainties for the asymmetry calculation.}
\begin{tabular}{ l | c c c}
\hline
\hline
& \multicolumn{3}{c}{Systematic error($\%$)} \\
\hline
Source & & ${\cal A}_{DK}$ & ${\cal A}_{D\pi}$ \\
\hline
Yield extraction & & 4.8 & 4.9 \\
Intrinsic detector charge asym & & 2.5 & 2.5\\
PID efficiency of prompt kaons & & 1.0 & $-$ \\
\hline
Total & & 5.5 & 5.5 \\
\hline
\hline
\end{tabular}
\label{tab:acpsys}
\end{table}

{\begin{center}
\section{\boldmath Summary}
\end{center}}
Using 274 million $B\bar B$ pairs collected with the Belle detector, we report studies of the suppressed decay  $B^{-}\to D_{cs}h^{-}$($h=K,\pi$). We observe $B^{-}\to D_{cs}\pi^{-}$ for the first time, with a significance of $5.8 \sigma$. The size of the signal is consistent with expectation based on measured branching fractions~\cite{pdg}. The significance for $B^{-}\to D_{cs}K^{-}$ is $2.7\sigma$ and we set an upper limit on the ratio of $B$ decay amplitudes $r_{B}$. This result is consistent with the measurement of  $r_{B}$ in the decay $B^{-} \to DK^{-}$, $D \to K_{S}\pi^{+}\pi^{-}$~\cite{da}.

{\begin{center}
\section*{Acknowledgments}
\end{center}}
We thank the KEKB group for the excellent operation of the
accelerator, the KEK Cryogenics group for the efficient
operation of the solenoid, and the KEK computer group and
the National Institute of Informatics for valuable computing
and Super-SINET network support. We acknowledge support from
the Ministry of Education, Culture, Sports, Science, and
Technology of Japan and the Japan Society for the Promotion
of Science; the Australian Research Council and the
Australian Department of Education, Science and Training;
the National Science Foundation of China under contract
No.~10175071; the Department of Science and Technology of
India; the BK21 program of the Ministry of Education of
Korea and the CHEP SRC program of the Korea Science and
Engineering Foundation; the Polish State Committee for
Scientific Research under contract No.~2P03B 01324; the
Ministry of Science and Technology of the Russian
Federation; the Ministry of Education, Science and Sport of
the Republic of Slovenia; the National Science Council and
the Ministry of Education of Taiwan; and the U.S.\
Department of Energy.


\end{document}